\documentclass{article}
\usepackage{amsmath,amsfonts,amssymb}
\usepackage{txfonts}
\usepackage[T1]{fontenc}
\begin{document}
\def\bd{\begin{displaymath}}\def\ed{\end{displaymath}}
\def\be{\begin{equation}}\def\ee{\end{equation}}
\def\bea{\begin{eqnarray}}\def\eea{\end{eqnarray}}
\def\ba{\begin{array}}\def\ea{\end{array}}
\def\nn{\nonumber}\def\lb{\label}\def\bb{\bibitem}

\def\a{\alpha}\def\b{\beta}\def\c{\chi}\def\d{\delta}\def\e{\epsilon}
\def\f{\phi}\def\g{\gamma}\def\h{\theta}\def\i{\iota}\def\j{\vartheta}
\def\k{\kappa}\def\l{\lambda}\def\m{\mu}\def\n{\nu}\def\o{\omega}\def
\p{\pi}\def\q{\psi}\def\r{\rho}\def\s{\sigma}\def\t{\tau}\def\u{\upsilon}
\def\V{\varphi}\def\w{\varpi}\def\y{\eta}\def\x{\xi}\def\z{\zeta}
\def\E{\varepsilon}

\def\D{\Delta}\def\F{\Phi}\def\G{\Gamma}\def\H{\Theta}\def\L{\Lambda}
\def\O{\Omega}\def\P{\Pi}\def\Q{\Psi}\def\S{\Sigma}\def\U{\Upsilon}\def\X{\Xi}

\def\lie{{\cal L}}\def\de{\partial}\def\na{\nabla}\def\per{\times}
\def\inf{\infty}\def\id{\equiv}\def\mo{{-1}}\def\ha{{1\over 2}}
\def\qu{{1\over 4}}\def\pro{\propto}\def\app{\approx}
\def\we{\wedge}\def\di{{\rm d}}\def\Di{{\rm D}}
\def\lra{\leftrightarrow}\def\bdot{\!\cdot\!}

\def\Ei{{\rm Ei}}\def\li{{\rm li}}\def\const{{\rm const}}\def\ex{{\rm e}}
\def\arcsh{{\rm arcsinh}}\def\arcch{{\rm arccosh}}
\def\arcth{{\rm arctanh}}\def\arccth{{\rm arccoth}}
\def\diag{{\rm diag}}

\def\gmn{g_{\m\n}}\def\ep{\e_{\m\n}}\def\ghmn{\hat g_{\m\n}}\def\mn{{\mu\nu}}
\def\dix{\int d^2x\ \sqrt{-g}\ }\def\ds{ds^2=}\def\sg{\sqrt{-g}}
\def\dhx{\int d^2x\ \sqrt{-\hat g}\ }\def\dex{\int d^2x\ e\ }
\def\sn{\mathop{\rm sn}\nolimits}\def\cn{\mathop{\rm cn}\nolimits}
\def\dn{\mathop{\rm dn}\nolimits}

\def\tors#1#2#3{T_{#1#2#3}}\def\curv#1#2#3#4{R_{#1#2#3#4}}
\def\af{asymptotically flat }\def\hd{higher derivative }\def\st{spacetime }
\def\fe{field equations }\def\bh{black hole }\def\as{asymptotically }
\def\eqs{equations }\def\eom{equations of motion }\def\trans{transformations }
\def\tran{transformation }\def\ther{thermodynamical }\def\coo{coordinates }
\def\bg{background }\def\gs{ground state }\def\bhs{black holes }
\def\sc{semiclassical }\def\hr{Hawking radiation }\def\sing{singularity }
\def\ct{conformal transformation }\def\cc{coupling constant }
\def\crel{commutation relations }\def\tl{transformation law }
\def\ns{naked singularity }\def\gi{gravitational instanton }
\def\rep{representation }\def\gt{gauge transformation }
\def\cco{cosmological constant }\def\em{electromagnetic }
\def\ssy{spherically symmetric }\def\cf{conformally flat }
\def\cur{curvature }\def\tor{torsion }\def\ms{maximally symmetric }
\def\coot{coordinate transformation }\def\diff{diffeomorphisms }
\def\gct{general coordinate transformations }\def\gts{gauge transformations }
\def\pb{Poisson brackets }\def\db{Dirac brackets }\def\ham{Hamiltonian }
\def\cd{covariant derivative }\def\dof{degrees of freedom }
\def\hdim{higher dimensional }\def\ldim{lower dimensional }
\def\SR{special relativity }
\def\dys{dynamical system }\def\cps{critical points }\def\dim{dimensional }
\def\sch{Schwarzschild }\def\min{Minkowski }\def\ads{anti-de Sitter }
\def\RN{Reissner-Nordstr\"om }\def\RC{Riemann-Cartan }\def\poi{Poincar\'e }
\def\KK{Kaluza-Klein }\def\pds{pro-de Sitter }\def\des{de Sitter }
\def\BR{Bertotti-Robinson }\def\MP{Majumdar-Papapetrou }
\def\GR{general relativity }\def\GB{Gauss-Bonnet }\def\CS{Chern-Simons }
\def\EH{Einstein-Hilbert }\def\EPG{extended \poi group }
\def\JT{Jackiw-Teitelboim }\def \schr{Schr\"odinger }
\def\dpa{deformed \poi algebra }\def\psm{Poisson sigma model }
\def\td{two-dimensional }\def\trd{three-dimensional }
\def\lt{Lorentz transformations }\def\com{center of mass }
\def\ab{asymptotic behavior}
\def\cor{commutation relations }\def\up{uncertainty principle }
\def\ev{expectation value }\def\bc{boundary conditions }
\def\tran{transformation }\def\ie{i.e.\ }

\def\PL#1{Phys.\ Lett.\ {\bf#1}}\def\CMP#1{Commun.\ Math.\ Phys.\ {\bf#1}}
\def\PRL#1{Phys.\ Rev.\ Lett.\ {\bf#1}}\def\AP#1#2{Ann.\ Phys.\ (#1) {\bf#2}}
\def\PR#1{Phys.\ Rev.\ {\bf#1}}\def\CQG#1{Class.\ Quantum Grav.\ {\bf#1}}
\def\NP#1{Nucl.\ Phys.\ {\bf#1}}\def\GRG#1{Gen.\ Relativ.\ Grav.\ {\bf#1}}
\def\JMP#1{J.\ Math.\ Phys.\ {\bf#1}}\def\PTP#1{Prog.\ Theor.\ Phys.\ {\bf#1}}
\def\PRS#1{Proc.\ R. Soc.\ Lond.\ {\bf#1}}\def\NC#1{Nuovo Cimento {\bf#1}}
\def\JoP#1{J.\ Phys.\ {\bf#1}} \def\IJMP#1{Int.\ J. Mod.\ Phys.\ {\bf #1}}
\def\MPL#1{Mod.\ Phys.\ Lett.\ {\bf #1}} \def\EL#1{Europhys.\ Lett.\ {\bf #1}}
\def\AIHP#1{Ann.\ Inst.\ H. Poincar\'e {\bf#1}}\def\PRep#1{Phys.\ Rep.\ {\bf#1}}
\def\AoP#1{Ann.\ Phys.\ {\bf#1}}\def\AoM#1{Ann.\ Math.\ {\bf#1}}
\def\JHEP#1{JHEP\ {\bf#1}}\def\JCAP#1{JCAP\ {\bf#1}}
\def\RMP#1{Rev.\ Mod.\ Phys.\ {\bf#1}}\def\AdP#1{Annalen Phys.\ {\bf#1}}
\def\grq#1{{\tt gr-qc/#1}}\def\hep#1{{\tt hep-th/#1}}\def\arx#1{{\tt arXiv:#1}}
\def\EPJ#1{Eur.\ Phys.\ J.\ {\bf#1}}

\def\rhd{\triangleright}\def\om{{\sqrt{\b k^2}}}\def\til{\tilde}
\def\rad{\sqrt{1+\b k^2}}\def\tD{{\tilde D}}
\def\xb{\bar x}\def\hx{\bar x}\def\f{\varphi}
\def\cH{{\cal H}}\def\cA{{\cal A}}\def\cD{{\cal D}}\def\cP{{\cal P}}\def\cC{{\cal C}}\def\cF{{\cal F}}\def\cI{{\cal I}}
\def\cJ{{\cal J}}\def\cT{{\cal T}}\def\cK{{\cal K}}\def\cG{{\cal G}}\def\cQ{{\cal Q}}\def\cS{{\cal S}}\def\cL{{\cal L}}
\def\rhd{\triangleright}\def\ot{\otimes}\def\op{\oplus}\def\Ex{{\rm exp}}\def\cO{{\cal O}}
\def\act{\rhd1}\def\bp{\bar\phi}
\begin{titlepage}
\pagenumbering{gobble}
\title{Quantum field theory in generalised Snyder spaces}
\vskip80pt
\author{S. Meljanac, D. Meljanac,\\
\small{Rudjer Bo\v skovi\'c Institute, Bijeni\v cka cesta 54, 10002 Zagreb, Croatia,}\\
\\
S. Mignemi and R. \v Strajn\\
\small{Dipartimento di Matematica e Informatica, Universit\`a di Cagliari,}\\
\small{viale Merello 92, 09123 Cagliari, Italy,}\\
\small{and INFN, Sezione di Cagliari, Cittadella Universitaria, 09042 Monserrato, Italy}}
\date{}
\maketitle
\vskip60pt
\begin{abstract}
We discuss the generalisation of the Snyder model that includes all possible
deformations of the Heisenberg algebra compatible with Lorentz invariance
and investigate its properties.
We calculate peturbatively the law of addition of momenta and the star product in the general case.
We also undertake the construction of a scalar field theory on these noncommutative spaces
showing that the free theory is equivalent to the commutative one, like in other models of
noncommutative QFT.
\end{abstract}

\end{titlepage}
\pagenumbering{arabic}

\section{Introduction}
Snyder spacetime \cite{Snyder} was introduced in 1947 as an attempt to avoid UV divergences in QFT.
By assuming a noncommutative structure of spacetime, and hence a deformation of the Heisenberg algebra,
it was possible to define a discrete
spacetime without breaking the Lorentz invariance, opening the possibility to smoothen the short-distance
behavior of quantum field theory.

The proposal was then forgotten for many years, until more recent times, when noncommutative
geometry has become an important field of research \cite{DFR}.
New models have been introduced, in particular the Moyal plane \cite{Moyal} and $\k$-Minkowski
geometry \cite{kappa}, and the formalism of Hopf algebras has been applied to their study \cite{Majid}.
However, contrary to Snyder's, the new models either break or deform the Lorentz group action on phase space.

In spite of the renewed interest in  spacetime noncommutativity and of its preservation of spacetime symmetries,
relatively few
investigations have been dedicated to the original proposal of Snyder from the point of view of noncommutative
geometry, except for a series of papers \cite{BM1}-\cite{kappaS}, where the model was extended to include more
general Lorentz-invariant models, as the one proposed by Maggiore \cite{Mag}, and the star product, coproduct and
antipodes of its Hopf algebra were calculated.
The model was also investigated in \cite {GL}, where it was considered from a geometrical point of view as a coset
in momentum space, with results equivalent to those of refs.\ \cite{BM1,BM2}.
Also the construction of QFT on Snyder spacetime was undertaken in these papers.

However, some basic properties of the Hopf algebra formalism for Snyder spaces have not yet been investigated:
for example the twist  has not been explicitly calculated.
Also the investigation of QFT has only been sketched. In \cite{BM2} a scalar field theory was
defined in terms of the star product, but no explicit calculation was carried on. Moreover, a non-Hermitian
representation was used, that complicates the formalism. Other approaches to Snyder QFT are based on a
five-dimensional formalism \cite{BSE,GL} .

Several efforts have also been devoted to the study of the classical and quantum aspects of the model from a
phenomenological point of view, without resorting to the formalism of noncommutative geometry,
especially in the nonrelativistic 3D limit \cite{CM}-\cite{LS}.
The most interesting results in this context are the clarification of its lattice-like properties, leading to deformed
uncertainty relations, and the study of the corrections induced on the energy spectrum of some simple physical systems.

In this letter, we extend previous investigations on the noncommutative geometry of the Snyder model in two
directions: first, we further generalise the model to include in the defining \cor all the terms compatible with
undeformed
Lorentz invariance. Among these generalisations, some have peculiar properties, for example it is possible to
construct models that describe a commutative spacetime, but nevertheless display nontrivial \cor between positions
and momenta, leading to deformed addition rules for momenta and nonlocal behavior in field theory.

Moreover, we improve the results of ref.\ \cite{BM2} on QFT, adopting a hermitian representation of the noncommutative
\coo and showing that with this definition the free field theory of a scalar particle is equivalent to the one in
noncommutative spacetime, similarly to other well-known models \cite{Sz,MS}.

\section{Snyder space and its generalisation}
We define generalised Snyder space as a deformation of ordinary phase space, generated by noncommutative \coo $\xb_\m$
and momenta $p_\m$ that span a deformed Heisenberg algebra $\bar\cH(\bar x,p)$,
\be\lb{snal}
[\xb_\m,\xb_\n]=i\b M_\mn\,\q(\b p^2),\qquad[p_\m,p_\n]=0,\qquad[p_\m,\xb_\n]=-i\f_\mn(\b p^2),
\ee
together with Lorentz generators $M_\mn$ that satisfy the standard relations
\bd
[M_\mn,M_{\r\s}]=i\big(\y_{\m\r}M_{\n\s}-\y_{\m\s}M_{\n\r}+\y_{\n\r}M_{\m\s}-\y_{\n\s}M_{\m\r}\big),
\ed
\be\lb{snal2}
[M_\mn,p_\l]=i\left(\y_{\m\l}p_\n-\y_{\l\n}p_\m\right),\qquad[M_\mn,\xb_\l]=i\left(\y_{\m\l}\xb_\n-\y_{\n\l}\xb_\m\right),
\ee
where the functions $\q(\b p^2)$ and $\f_\mn(\b p^2)$ are constrained so that the Jacobi identities hold,
$\b$ is a constant of the order of $1/M_{Pl}^2$, and $\y_\mn=$\ diag $(-1,1,1,1)$.
The \cor (\ref{snal})-(\ref{snal2}) generalise those of the Snyder spaces originally investigated in \cite{BM1,BM2},
that are recovered for $\q=$ const. Special cases are the Snyder realisation \cite{Snyder}, and the Maggiore realisation
\cite{Mag}.

We recall that in its undeformed version, the Heisenberg algebra $\cH(x,p)$ is generated by commutative
\coo $x_\m$ and momenta $p_\m$, satisfying
\be
[x_\m,x_\n]=[p_\m,p_\n]=0,\qquad[p_\m,x_\n]=-i\y_\mn.
\ee
The action of $x_\m$ and $p_\m$ on functions $f(x)$ belonging to the enveloping algebra $\cA$ generated by the $x_\m$
is defined as
\be
x_\m\rhd f(x)=x_\m f(x),\qquad p^\m\rhd f(x)=-i{\de f(x)\over \de x_\m}.
\ee

The noncommutative \coo $\xb_\m$ and the Lorentz generators $M_\mn$ in \eqref{snal}-\eqref{snal2} can be expressed in terms of
commutative \coo $x_\m$ and momenta $p_\m$ as \cite{BM1,BM2}
\be\lb{real}
\xb_\m=x_\m \f_1(\b p^2)+\b \,x\bdot p\,p_\m\f_2(\b p^2)+\b p_\m\c(\b p^2),
\ee
\be
M_\mn=x_\m p_\n-x_\n p_\m.
\ee
Notice that the function $\c$ does not appear in the defining relations \eqref{snal}-\eqref{snal2}, but takes into account ambiguities
arising from operator ordering of $x_\mu$ and $p_\mu$ in equation \eqref{real}.

In terms of the realisation \eqref{real}, the functions $\f_\mn$ in (\ref{snal}) read
\be\lb{phi}
\f_\mn=\y_\mn\f_1+\b p_\m p_\n\f_2,
\ee
while the Jacobi identities are satisfied if
\be\lb{jac}
\q =-2\f_1\f'_1+\f_1\f_2-2\b p^2\f_1'\f_2,
\ee
where the prime denotes a derivative with respect to $\b p^2$. In particular, the function $\psi$
does not depend on the function $\chi$.

Inserting \eqref{jac} into \eqref{snal}, it is easy to check that the \coo $\xb_\m$ are commutative for
$\f_2={2\f_1'\f_1\over\f_1-2\b p^2\f_1'}$,
and correspond to Snyder spaces for $\f_2={1+2\f_1'\f_1\over\f_1-2\b p^2\f_1'}$.
In particular, the Snyder realisation \cite{Snyder} is recovered for $\f_1=\f_2=1$, and the Maggiore realisation
\cite{Mag} for $\f_1=\sqrt{1-\b p^2}$, $\f_2=0$. Another interesting exact realisation of generalised
Snyder spaces is obtained for $\q=s=$ const, and reads
\be \label{conf}
\xb_\m=x_\m+{\b s\over4}\,K_\m,
\ee
where $K_\m=x_\m p^2-2x\bdot p\,p_\m\ $ are the generators of special conformal transformations in momentum space,
satisfying $[K_\m,K_\n]=0$.

The algebra (\ref{snal})-(\ref{snal2}) includes as special cases both commutative spaces, $\q=0$, and Snyder spaces, $\q=1$.
Since the Lorentz transformations  are not deformed, its Casimir operator is the same as for the \poi algebra, $\cC=p^2$.

\section{Coproduct and star product}
The generalised Snyder spaces defined above can be investigated using the Hopf-algebra formalism developed
in refs.\ \cite{Svrtan}-\cite{KMSS} and shortly reviewed in \cite{MM}, to which we refer for more details\footnote{A more
rigorous treatment, including the full phase space, is given by the Hopf algebroid formalism \cite{JMS}. However, we shall
not need it in the following.}. In this way, one
can deduce the properties of the algebra associated to Snyder space starting from its realisation \eqref{real}.

It can be shown that for a general Hopf algebra $\cA$ \cite{Svrtan}-\cite{KMSS},
\be\lb{defP}
e^{ik\cdot\xb}\rhd e^{iq\cdot x}=e^{i\cP(k,q)\cdot x+i\cQ(k,q)},
\ee
and
\be\lb{defK}
e^{ik\cdot\xb}\rhd1=e^{i\cK(k)\cdot x+i\cL(k)},
\ee
where  eqs.\ \eqref{defP} and \eqref{defK} can be seen as the defining
relations for the functions $\cP$, $\cQ$, $\cK$ and $\cL$.

It is easily seen that $\cP_\m(\l k,q)$ satisfies the differential equation \cite{Svrtan}-\cite{KMSS}
\be\lb{diffe}
{d\cP_\m(\l k,q)\over d\l}=k_\a\f_\m{}^{\,\a}\Big(\cP(\l k,q)\Big),
\ee
where $\l$ is a real parameter and
\be
\cP_\m(k,0)=\cK_\m(k),\qquad\cP_\m(0,q)=q_\m.
\ee

Analogously, it can be shown that $\cQ_\m(\l k,q)$ satisfies the differential equation
\be\lb{Q}
{d\cQ(\l k,q)\over d\l}=k_\a\c^\a\Big(\cP(\l k,q)\Big),
\ee
with $\c^\a\id\b p^\a\c(\b p^2)$
and
\be
\cQ(k,0)=\cL(k),\qquad\cQ(0,q)=0.
\ee

The generalised addition of momenta $k_\m$ and $q_\m$ is then defined as
\be
k_\m\oplus q_\m=\cD_\m(k,q),\qquad{\rm with}\quad\cD_\m(k,0)=k_\m,\quad\cD_\m(0,q)=q_\m,
\ee
where the function $\cD_\m(k,q)$ can be obtained from $\cP_\m(k,q)$ as \cite{BM2,kappaS,JMP}
\be
\cD_\m(k,q)=\cP_\m(\cK^\mo(k),q),
\ee
and the function $\cK^\mo_\m(k)$ is the inverse map of $\cK(k)$, \ie $\cK^\mo_\m(\cK(k))=k_\m$.
Remarkably, from \eqref{diffe} and \eqref{phi} it follows that $\cP_\m(k,q)$ and hence $\cD_\m(k,q)$ do not depend
on the function $\c$ in \eqref{real}.

From \eqref{defK}, it is also possible to calculate the star product of two plane waves, that turns out to be
\be\lb{star}
e^{ik\cdot x}\star e^{iq\cdot x} = e^{i\cD(k,q)\cdot x+i\cG(k,q)},
\ee
where
\be\lb{G}
\cG(k,q)=\cQ(\cK^\mo(k),q)-\cQ(\cK^\mo(k),0).
\ee
Note that $\cG$ vanishes if $\c=0$.

Finally, the coproduct for the momenta $\D p_\m$ can be written as usual in terms of $\cD_\m(k,q)$ as
\be\lb{cop}
\D p_\m=\cD_\m(p\ot1,1\ot p).
\ee
The previous definitions imply that the addition of momenta and the coproduct do not depend on $\c$.
Following the steps sketched above, the coproducts of momenta were found for special cases in ref.\ \cite{BM2}:
for example, for the Snyder realisation \cite{Snyder},
\begin{equation} \label{coprod}
\Delta p_{\mu} = \frac{1}{1-\b p_{\alpha}\otimes p^{\alpha}} \left(p_{\mu} \otimes 1 - \frac{\b}
{1+\sqrt{1+\b p^2}}\, p_{\mu}p_{\alpha} \otimes p^{\alpha} + \sqrt{1+\b p^2} \otimes p_{\mu} \right).
\end{equation}

Finally, we recall that the antipodes for Snyder space are trivial \cite{BM2},
\be
S(p_\m)=-p_\m,\qquad S(M_\mn)=-M_\mn.
\ee

\section{First order expansion}
The study of the Hopf algebra for the generalised Snyder model is difficult, but can be tackled using a perturbative
approach, by expanding the realisation \eqref{real} of the noncommutative \coo in powers of $\b$.
The expansion gives
\be \label{1redx}
\xb_\m=x_\m+\b\,(s_1x_\m p^2+s_2x\bdot p\,p_\m+cp_\m)+\cO(\b^2),
\ee
with independent parameters $s_1$, $s_2$, $c$.
The \cor do not depend on the parameter $c$ and to first order are given by
\be\lb{cor}
[\xb_\m,\xb_\n]=i\b sM_\mn +o(\beta^2),\qquad[p_\m,\xb_\n]=-i\,\big[\y_\mn(1+\b s_1p^2)+\b s_2p_\m p_\n\big]+\cO(\beta^2),
\ee
where $s=s_2-2s_1$.

The models of ref.\  \cite{BM1,BM2} are recovered for $s_2=1+2s_1$. Moreover,
for $s_1=0$, $s_2=1$, eqs.\ \eqref{1redx}-\eqref{cor} reproduce the exact Snyder realisation, while for $s_1=-\ha$, $s_2=0$
they give the first-order expansion of the Maggiore realisation. For $s_2=2s_1$, spacetime is commutative to first order
in $\beta$, while for $s_1=-s/4$, $s_2=s/2$, $c=0$ one gets the exact realisation \eqref{conf}.

From \eqref{diffe} one can calculate the first order expression for the function $\mathcal{P}_\mu (k,q)$ in the general
case, which reads
\begin{eqnarray}\lb{Plin}
\mathcal{P}_\mu (k,q)&=&
 k_\mu +q_\mu + \beta\, \Bigg[ \left(s_1 q^2 + \left( s_1+\frac{s_2}{2} \right) k\cdot q +
\frac{s_1+s_2}{3} k^2 \right) k_\mu\cr \nonumber\\
&&+ s_2 \left( k\cdot q +\frac{k^2}{2} \right) q_\mu \Bigg]+\cO(\beta^2),
\end{eqnarray}
from where it follows that
\begin{equation}
\cK^{-1}_\mu (k) =k_\mu -\frac{\beta}{3}(s_1+s_2) k^2 k_\mu +\cO(\beta^2).
\end{equation}

These results allow one to write down the generalised addition law of the momenta $k_\m$ and $q_\m$ to first order
\be\lb{linadd}
(k\op q)_\m= \mathcal{D}_\mu (k,q)= k_\m+q_\m+\b\left[s_2k\bdot q\,q_\m+s_1q^2k_\m
+\left(s_1+{s_2\over2}\right)k\bdot q\,k_\m+{s_2\over2}k^2q_\m\right]+\cO(\b^2).
\ee
It is interesting to remark that for $s_2=2s_1\neq0$, although spacetime is commutative up to the first order in $\beta$,
the addition of momenta is deformed,
\be
(k\op q)_\m\ne k_\m+q_\m.
\ee

The Lorentz transformations of momenta are instead not deformed, and denoting them by $\L(\x,p)$, with $\x$ the rapidity
parameter, the law of addition of momenta implies that
\be
\L(\x,k\op q)=\L(\x_1,k)\op\L(\x_2,q)
\ee
is satisfied for $\x_1=\x_2=\x$. Hence there are no backreaction factors in the sense of ref. \cite{GM,maj}.
This means that in composite systems the boosted momenta of the single particles are independent of the momenta of the
other particles in the system. This confirms the results obtained from general arguments in \cite{IMS} and \cite{MM}.

The coproduct to first order can be read from (\ref{linadd}) and is given by
\begin{equation} \label{coprod1ord}
\Delta p_\mu = \Delta_0 p_\mu + \beta \left[ s_1 p_\mu \otimes p^2 + s_2 p_\alpha \otimes p^\alpha p_\mu +\left( s_1 +
\frac{s_2}{2} \right) p_\mu p_\alpha \otimes p^\alpha + \frac{s_2}{2} p^2 \otimes p_\mu \right] + \cO(\beta^2).
\end{equation}

\section{Field theory for the Snyder realisation}

The scalar field theory on Snyder spacetime was investigated  in \cite{BM2} using the Snyder realisation \cite{Snyder},
\begin{equation}\lb{snreal}
\xb_{\mu}=x_{\mu}+\b\,x\bdot p\,p_{\mu}.
\end{equation}
However, this realisation is not Hermitian, and hence also the resulting action functional is not Hermitian.
This causes some problems, in particular in the definition of a measure for the integral, while,
as we shall show, with a Hermitian realisation the free field action reduces to the usual commutative form.
Similar consideration have been made in \cite{MS} for the $\k$-Minkowski case.

The realisation \eqref{snreal} can be made Hermitian by adding its adjoint, yielding
\be\lb{herreal}
\hx^\m=x^\m+{\b\over2}\left(x\bdot p\, p^\m+p^\m\,p\bdot x\right)=x^\m+\b\,x\bdot p\,p^\m-{5i\over2}\b\, p^\m,
\ee
where we have used the canonical \cor to rearrange the expression.

For the Snyder realisation, one obtains \cite{BM2}
\be
\cP_\m(k,q)={q_\m+\left[{\sin\om\over\om}+{k\cdot q\over k^2}\left(\cos\om-1\right)\right]k_\m\over\cos\om-
{k\cdot q\over k^2}\om\,\sin\om},
\ee
and
\be
\cK_\m(k)={\tan\om\over\om}\,k_\m,
\ee
from which it follows that
\be\lb{cD}
\cD_\m(k,q)={1\over1-\b k\bdot q}\left[\left(1-{\b\,k\bdot q\over1+\rad}\right)k_\m+\rad\, q_\m\right],
\ee

Using eqs.\ \eqref{Q} and \eqref{G}, one can compute the functions $\cQ$ and $\cG$ for the hermitian realisation
\eqref{herreal}, obtaining
\be
\cQ(k,q)={5i\over2}\ln\left[\cos\om-{k\bdot q\over k^2}\om\,\sin\om\right],
\ee
and
\be
\cG(k,q)={5i\over2}\ln\left[1-\b\,k\bdot q\right].
\ee

According to \eqref{star}, the star product for plane waves in the hermitian realisation \eqref{herreal} is therefore
\be
e^{ik\cdot x}\star e^{iq\cdot x}={e^{i\cD(k,q)\cdot x}\over(1-\b\,k\bdot q)^{5/2}},
\ee
with $\cD_\m(k,q)$ given by \eqref{cD}.

The action of a noncommutative scalar field $\bar\phi(\xb)$ on the identity (ground state) of $\cA$ is defined as
\be
\bp(\xb)\act=\phi(x),\qquad\bp(\xb)^2\act=(\phi\star\phi)(x).
\ee
The action functional for a noninteracting massive real scalar field can then be defined as
\be
S[\phi]=\ha\int d^4\xb\,(\de_\m\bp\,\de^\m\bp+m^2\bp^2)\act=\ha\int d^4x\,(\de_\m\phi\star\de^\m\phi+m^2\phi\star\phi)
\ee

To write the action in simpler form, we compute the star product of two real scalar fields
$\phi(x)$ and $\q(x)$ by expanding in Fourier series,
\be
\phi(x)=\int d^4k\,\til\phi(k)e^{ikx}.
\ee
Then
$$\int d^4x\ \q(x)\star\phi(x)=\int d^4x\int d^4k\,d^4q\ \tilde\q(k)\,
\tilde\phi(q)\ e^{ik\cdot x}\star e^{iq\cdot x}=$$
\be\lb{prod}
\int d^4k\,d^4q\ \til\q(k)\,\til\phi(q)\int d^4x\,{e^{i\cD(k,q)\cdot x}\over(1-\b\,k\bdot q)^{5/2}}=
\int d^4k\,d^4q\ \til\q(k)\,\til\phi(q)\ {\d^{(4)}\big(\cD(k,q)\big)\over(1-\b\,k\bdot q)^{5/2}}.
\ee
\smallskip

Now, since $\cD_\m(k,q)$ vanishes only for $q=-k$,
\be
\d^{(4)}\big(\cD(k,q)\big)={\d^{(4)}(q+k)\over\left|\det\left({\de\cD_\m(k,q)\over\de q_\n}\right)\right|_{q=-k}}.
\ee
On the other hand,
\be
{\de\cD_\m(k,q)\over\de q_\n}\Bigg|_{q=-k}={1\over1+\b k^2}\left(\rad\,\d_\m^\n-{\b k_\m k^\n\over1+\rad}\right),
\ee
and then
\be\lb{det}
\left|\det\left({\de\cD_\m(k,q)\over\de q_\n}\right)\right|_{q=-k}={1\over(1+\b k^2)^{5/2}}.
\ee
The term in \eqref{det} cancels with the one coming from the star product in \eqref{prod}, and finally one obtains
\be\lb{acx}
\int d^4x\ \q(x)\star\phi(x)=\int d^4x\ \q(x)\,\phi(x).
\ee
Hence, the integral of a star product of two fields in the Hermitian realisation can be reduced to the integral of an ordinary
product of commutative functions. The same happens in the Moyal \cite{Sz} and $\k$-Minkowski case \cite{MS}.
We conjecture that this is a universal property of noncommutative models in a Hermitian realisation, although we are not able
to prove it.

In particular, for a free scalar field, the action $S[\phi]$ can then be written as
\be\lb{action}
S[\phi]=\ha\int d^4x\left(\de_\m\phi\star\de_\m\phi+m^2\phi\star\phi\right)=\ha\int d^4x\left(\de_\m\phi\,\de_\m\phi+m^2\phi^2\right).
\ee

\section{Field theory for the linearised theory}
The previous calculations can be extended to the generalised Snyder models by a perturbative expansion in $\b$ up to first order.
This is a good approximation for small momenta, although cannot be trusted in the ultraviolet region.

For simplicity, we shall limit our consideration to the Snyder models of ref.\ \cite{BM2}, neglecting the generalisation of section
2. In this case $s_2=1+2s_1$, and the realisation \eqref{1redx} reduces to
\be\lb{lreal}
\xb^\m=x^\m(1+\b s_1p^2)+\b(1+2s_1)\,x\bdot p\,p^\m+\cO(\b^2).
\ee
The Snyder realisation discussed in the previous section is obtained for $s_1=0$. In general \eqref{lreal} is not Hermitian, but a
Hermitian realisation can be obtained as before by redefining
\be\lb{linher}
\xb^\m=x^\m(1+\b s_1p^2)+\b(1+2s_1)\,x\bdot p\,p^\m-i\b\left({5\over2}+6s_1\right)p^\m+\cO(\b^2).
\ee

For this realisation, eq.\ \eqref{Plin} reduces to
\bea
\cP_\m(k,q)&=&k_\m+q_\m+\b\left[s_1q^2+\left(\ha+2s_1\right)k\bdot q+\left({1\over3}+s_1\right)k^2\right]k_\m+\cr
&&\b(1+2s_1)\left[{k^2\over2}+k\bdot q\right]q_\m+\cO(\b^2),
\eea
and
\be
\cK_\m(k)=\left[1+\b\left({1\over3}+s_1\right)k^2\right]k_\m+\cO(\b^2),
\ee
from which it follows that
\be\lb{Dlin}
\cD_\m(k,q)=k_\m+q_\m+\b\left[s_1q^2+\left(\ha+2s_1\right)k\bdot q\right]k_\m+\b(1+2s_1)\left[{k^2\over2}+k\bdot q\right]q_\m+\cO(\b^2).
\ee

One can now compute the linearised $\cQ$ and $\cG$ from \eqref{Q} and \eqref{G}, obtaining
\be
\cQ(k,q)=-i\b\left({5\over2}+6s_1\right)\left({k^2\over2}+k\bdot q\right)+\cO(\b^2),
\ee
and
\be
\cG(k,q)=-i\b\left({5\over2}+6s_1\right)k\bdot q+\cO(\b^2).
\ee
The star product for the Hermitian realisation \eqref{linher} is therefore
\be\lb{linstar}
e^{ik\cdot x}\star e^{iq\cdot x}=\left[1+\b\left({5\over2}+6s_1\right)k\bdot q\right]e^{i\cD(k,q)\cdot x}+\cO(\b^2).
\ee
where $\cD$ is given in eq.\ \eqref{Dlin}.

The product of fields can now be computed as before expanding in noncommutative plane waves, and gives
\be
\int d^4x\ \q(x)\star\phi(x)=\int d^4k\,d^4q\ \til\q(k)\,\til\phi(q)\left[1+\b\left({5\over2}+6s_1\right)k\bdot q+\cO(\b^2)\right]
\d^{(4)}\big(\cD(k,q)\big).
\ee
Now, it is easy to see that $\cD(k,q)$ vanishes only for $q_\m=-k_\m$, and then
\be
\d^{(4)}\big(\cD(k,q)\big)={\d^{(4)}(q+k)\over\left|\det\left({\d\cD_\m(k,q)\over\de q_\n}\right)\right|_{q=-k}}.
\ee
On the other hand,
\be
{\de\cD_\m(k,q)\over\de q_\n}\Big|_{q=-k}=\d_\m^\n-\b\left[\left(\ha+s_1\right)k^2\d_\m^\n+\left(\ha+2s_1\right)k_\m k^\n\right]+\cO(\b^2),
\ee
and hence
\be\lb{lindet}
\left|\det\left({\de\cD_\m(k,q)\over\de q_\n}\right)\right|_{q=-k}=1-\b\left({5\over2}+6s_1\right)k^2+\cO(\b^2).
\ee
Again the corrections coming from \eqref{linstar} and \eqref{lindet} cancel and one recovers \eqref{acx},
showing that also for general realisations at the linear level the integral of the star product in a Hermitian realisation
is equivalent to the ordinary product.
In particular, eq.\ \eqref{action} still holds and the free field propagator coincides with the commutative one.

Interaction terms can be added to the action through the star product; for example, a cubic interaction can be described by
\be\lb{int}
I^{(3)}=\int d^4x\ \phi\star(\phi\star\phi).
\ee
Notice that because of the nonassociativity of the star product the ordering of the products is important; one may also define
the interaction using symmetrised forms of \eqref{int}.

The computation of the star products in \eqref{int} involves the evaluation of the vertex operator
$\cD^{(3)}_\m(k_1,k_2,k_3)=\cD_\m(k_3,\cD(k_2,\cD(k_1)))$. This can be easily done using the results of the previous sections.
We leave the calculation of these terms and of loop corrections to future investigations.

\section{Conclusions}
In this paper we have extended the study of the Snyder model to its most general realisations compatible with undeformed
Lorentz invariance.
Some of the new models have peculiar properties, for example they can display commutative spacetime geometry, but nontrivial
Heisenberg algebra, obeying nonstandard laws for the addition of momenta.

We have also constructed a scalar QFT on Snyder spaces along the lines of \cite{BM2}. The main improvement of our approach has
been the use of a Hermitian representation for the noncommutative spacetime coordinates. This has allowed us to show that the
noninteracting terms in the action can be reduced to the ordinary noncommutative form, as in other noncommutative models
\cite{Sz,MS}.
This might be a universal property of noncommutative models with Hermitian action, and it would be interesting to further
investigate the origin of this property. The interacting theory can also be studied in our formalism, and we plan to
pursue this topic in future work.

\section*{Acknowledgements}
The work of S. Meljanac has been supported by Croatian Science Foundation under the project IP-2014-09-9582,
as well as by the H2020 Twinning project No. 692194, "RBI-T-WINNING".
S. Mignemi wishes to thank the Rudjer Bo\v skovi\'c Institute for hospitality during the preparation of this work.

\end{document}